\documentclass[english,aip,superscriptaddress, reprint]{revtex4-1}
\usepackage[T1]{fontenc}
\usepackage[latin9]{inputenc}
\usepackage{booktabs}
\usepackage{amsmath}
\usepackage{graphicx}
\usepackage{wasysym}
\usepackage{color}

\makeatletter

\providecommand{\tabularnewline}{\\}

\usepackage{subfigure}
\usepackage{dcolumn}
\usepackage{bm}

\makeatother

\usepackage{babel}
\begin{document}

\title{Two-phase Thermodynamic Model for Computing Entropies of Liquids
Reanalyzed}

\author{Tao Sun}
\email{tsun@ucas.ac.cn}

\selectlanguage{english}%

\affiliation{Key Laboratory of Computational Geodynamics, College of Earth Sciences,
University of Chinese Academy of Sciences, 100049, Beijing, China}

\author{Jiawei Xian}

\affiliation{Institute of Applied Physics and Computational Mathematics, 100088,
Beijing, China}

\affiliation{Software Center for High Performance Numerical Simulation, China
Academy of Engineering Physics, Beijing 100088, China}

\author{Huai Zhang}

\affiliation{Key Laboratory of Computational Geodynamics, College of Earth Sciences,
University of Chinese Academy of Sciences, 100049, Beijing, China}

\author{Zhigang Zhang}

\affiliation{Key Laboratory of Earth and Planetary Physics, Institute of Geology
and Geophysics, Chinese Academy of Sciences, 100029, Beijing, China}

\author{Yigang Zhang}

\affiliation{Key Laboratory of Earth and Planetary Physics, Institute of Geology
and Geophysics, Chinese Academy of Sciences, 100029, Beijing, China}
\begin{abstract}
The two-phase thermodynamic (2PT) model {[}J. Chem. Phys., \textbf{119},
11792 (2003){]} provides a promising paradigm to efficiently determine
the ionic entropies of liquids from molecular dynamics (MD). In this
model, the vibrational density of states (VDoS) of a liquid is decomposed
into a diffusive gas-like component and a vibrational solid-like component.
By treating the diffusive component as hard sphere (HS) gas and the
vibrational component as harmonic oscillators, the ionic entropy of
the liquid is determined. Here we examine three issues crucial for
practical implementations of the 2PT model: (i) the mismatch between
the VDoS of the liquid system and that of the HS gas; (ii) the excess
entropy of the HS gas; (iii) the partition of the gas-like and solid-like
components. Some of these issues have not been addressed before, yet
they profoundly change the entropy predicted from the model. Based
on these findings, a revised 2PT formalism is proposed and successfully
tested in systems with Lennard-Jones potentials as well as many-atom
potentials of liquid metals. Aside from being capable of performing
quick entropy estimations for a wide range of systems, the formalism
also supports fine-tuning to accurately determine entropies at specific
thermal states. 
\end{abstract}
\maketitle

\section{introduction}

Entropy $(S)$ is a fundamental while somewhat peculiar thermodynamic
quantity in molecular dynamics (MD) simulations.\cite{allen1989,frenkel2002}
It is one of the basic inputs, along with energy $(E)$, pressure
$(P)$, density ($\rho$), and temperature $(T)$, to determine the
system's free energy essential for establishing phase diagrams and
other thermodynamic properties. However, unlike $E$, $P$, $\rho$,
and $T$, entropy can not be defined as time averages over phase space
trajectories,\cite{allen1989,frenkel2002} and special techniques
are needed to evaluate entropy in MD. For solids, entropy can be evaluated
through the power spectrum (vibrational density of states, VDoS) of
the velocity autocorrelation function (VACF) using phonon gas model
(PGM).\cite{wallace1972,grimvall1999,fultz2010,sun2010,zhang2014}
PGM takes into account lattice anharmonicity via temperature-dependent
phonon frequencies and is applicable even to strongly anharmonic crystals.\cite{sun2014,lu2017}
However, PGM is not suitable for liquids as diffusion of atoms, a
characteristic feature of fluids, cannot be described by phonons.
To overcome this difficulty, Lin et al.\cite{lin2003} proposed an
ingenious two-phase thermodynamic (2PT) model. In this model, the
VDoS of a liquid is decomposed into a diffusive gas-like component
and a vibrational solid-like component. Entropy associated with the
diffusive component is determined from the hard spheres (HS) model,
that with the vibrational component is from the harmonic oscillator
(phonon) model. The 2PT model requires only one post-processing of
the MD trajectory to evaluate entropy. Thus it is much more efficient
than the conventional thermodynamic integration (TI) approach which
involves many separate MD simulations along the integration path.
Because of this, the 2PT model has attracted considerable attention
and has been applied to many systems including Ar,\cite{lin2003}
CO$_{2}$,\cite{huang2011a} H$_{2}$O,\cite{lin2010} liquid metals,\cite{desjarlais2013,robert2015,jakse2016}
silicates and oxides,\cite{boates2013} etc.

In essence, the 2PT model relies on phonons to describe vibration
and HS to describe diffusion. The former is a natural extension of
the well-established PGM for solids and can be regarded as reliable.
The latter, as it turns out, is more problematic and requires careful
consideration. Here we focus on three issues that are related to HS.
The first issue, as noticed by Desjarlais,\cite{desjarlais2013} is
that the VDoS of HS gas declines more slowly with frequency than that
of the actual liquid. As a result, the VDoS of the gas-like component
is larger than the total VDoS at high frequencies. This mismatch in
VDoS was found to cause significant overestimation of entropy (up
to 0.3 to 0.4 $k_{B}$ per atom) in \emph{ab initio} MD (AIMD) simulations
of liquid metals.\cite{desjarlais2013} The second issue, which has
not been addressed so far, is the explicit formula used to evaluate
the excess entropy ($S_{{\rm ex}}$) of HS gas. The excess entropy
is defined as the entropy difference between non-ideal and ideal gases
under the same physical condition. Here the \emph{same} physical condition
refers to either identical $T$ and $\rho$, or identical $T$ and
$P$. In the original 2PT paper,\cite{lin2003} Lin et al. applied
the Carnahan-Starling formula\cite{carnahan1970} to compute $S_{{\rm ex}}$
under identical $T$ and $\rho$. However, a closer inspection would
reveal that the Carnahan-Starling formula actually corresponds
to $S_{{\rm ex}}$ under identical $T$ and $P$.\cite{carnahan1970,oconnell2005,hansen2006}
The two formulas, $S_{{\rm ex}}(T,\:\rho)$ and $S_{{\rm ex}}(T,\:P)$,
differ by a term $\ln(z)$, where $z$ is the compressibility of the
HS gas.\cite{oconnell2005} As will be shown in the paper, removing
this $\ln(z)$ term causes profound changes to the predicted entropies.
It is surprising that the predicability of a
model hinges on an apparently misplaced formula. Resolving this puzzle
leads to the third issue we would address: the partition of the gas-like
and solid-like components. Indeed, entropy predicted by the 2PT model
relies on the gas-solid partition and this partition can be refined
to improve the accuracy of the model. 

In this paper, we introduce a revised 2PT formalism which resolves
the above three issues. The formalism is first validated with liquid
Ar using Lennard-Jones potential, in the same fashion as Lin et al.'s
original work. It is then applied to liquid metals using Sutton-Chen
many-atom potentials.\cite{sutton1990} The adoption of classical
potentials, rather than AIMD, allows us to perform extensive TI calculations
to check the accuracy of the entropies from the 2PT model. The analytic
nature of Sutton-Chen potentials is also ideal for demonstrating how
the softness of interatomic potentials affects the predicted 2PT entropies.
We stress that while only classical potentials are considered in the
present paper, the proposed formalism should also be applicable to
AIMD simulations as the underlying physics is identical.

% This paper is organized as follows: Section 2 gives a brief summary of the original 2PT formalism. Using liquid Ar as the model system, we entropy change after imposing the right formula for $S_{\rm ex}$. 

\section{original 2pt formalism}

The starting point of a 2PT calculation\cite{lin2003} is the VACF
$C(t)$ and its power spectrum (VDoS) $F(\nu)$, evaluated in MD as 

\begin{align}
C(t) & =\frac{1}{3(N-1)}\sum_{i=1}^{N}m\left\langle \mathbf{v}_{i}\left(0\right)\cdot\mathbf{v}_{i}\left(t\right)\right\rangle ,\label{eq: VACF}
\end{align}

\begin{align}
F(\nu) & =\frac{12}{k_{B}T}\int_{0}^{\infty}C(t)\cos(2\pi\nu t)dt.\label{eq:DoS}
\end{align}
Here $m$ is the mass of an atom, $N$ is the total number of atoms,
${\rm {\bf v}}_{i}\left(t\right)$ denotes the velocity of the \emph{i}th
atom at time $t$, and $\left\langle \cdots\right\rangle $ stands
for ensemble average, $k_{B}$ is the Boltzmann's constant. With such
definitions, we have (i) $C(0)=k_{B}T$; (ii) $\int_{0}^{\infty}F(\nu)d\nu=3$;
(iii) the VDoS at zero frequency $F\left(0\right)$ is proportional
to the diffusion coefficient $D$ as\cite{hansen2006,mcquarrie2000}
\begin{align}
F\left(0\right) & =\frac{12m}{k_{B}T}D.\label{eq: f0 and d}
\end{align}
A non-zero $F(0)$ indicates that the system is in fluid state.

Next, $C(t)$ and $F\left(\nu\right)$ are partitioned into a gas-like
component and a solid-like component as 
\begin{align}
C(t) & =f_{g}C_{g}(t)+(1-f_{g})C_{s}(t),\label{eq: vdos decomposition}\\
F\left(\nu\right) & =f_{g}F_{g}\left(\nu\right)+\left(1-f_{g}\right)F_{s}\left(\nu\right),
\end{align}
where the subscripts distinguish the gas-like($g$) and solid-like($s$)
subsystems, respectively, $f_{g}$ denotes the gas-like fraction of
the system and takes value between $0$ (completely solid) and $1$
(completely gas). According to the Enskog theory of HS model, $C_{g}(t)$
equals $k_{B}T\exp(-\alpha t)$,\cite{mcquarrie2000,hansen2006} where
$\alpha$ is a parameter to be calculated, and
\begin{align}
F_{g}(\nu) & =\frac{12\alpha}{\alpha^{2}+4\pi^{2}\nu^{2}}.\label{eq:Fg_HS}
\end{align}
Note $F_{g}(\nu)$ obeys the sum rule $\int_{0}^{\infty}F_{g}(\nu)d\nu=3$,
an useful feature in subsequent entropy evaluations. Since the diffusive
VDoS at zero frequency $F(0)$ should be completely attributed to
the gas component, we have 
\begin{align}
F(0) & =f_{g}F_{g}(0)=\frac{12f_{g}}{\alpha}.\label{eq:alpha_HS}
\end{align}
Thus $\alpha$, and subsequently $F_{g}(\nu)$ , can be determined
once $f_{g}$ is known. The VDoS associated with the solid-like component
$(1-f_{g})F_{s}(\nu)$ can then be determined as $F(\nu)-f_{g}F_{g}(\nu)$. 

To evaluate $f_{g}$, Lin et al made two further assumptions:\cite{lin2003}
(i) $f_{g}=\frac{D}{D_{0}}$, where $D_{0}$ is the HS diffusivity
in the low density limit (the Chapman-Enskog result); (ii) the diffusivity
of the gas component can be determined analytically using the Enskog
theory for dense HS, and it is $1/f_{g}$ times larger than the diffusivity
of the whole system. With these two assumptions, $f_{g}$, as well
as the packing fraction of HS, are uniquely determined and the partition
of gas-solid components is accomplished (see Appendix A for a general
derivation) . 

The partition of gas-solid components can be interpreted as the liquid
system under study is dynamically equivalent to a combination of two
subsystems: one is HS gas with $f_{g}N$ particles, the other is harmonic
oscillators with VDoS equaling $N(1-f_{g})F_{s}(\nu)$. Entropy of
the liquid equals the sum of the entropies of subsystems. For the
solid subsystem, the associated entropy $S_{s}$ is determined as
\begin{align}
S_{s} & =Nk_{_{B}}\left(1-f_{g}\right)\int_{0}^{\infty}F_{s}\left(\nu\right)W_{s}\left(\nu\right)d\nu,\label{eq:entropy_solid}
\end{align}
where 
\begin{align}
W_{s}\left(\nu\right) & =\frac{h\nu/k_{B}T}{e^{h\nu/k_{B}T}-1}-\ln\left[1-e^{-h\nu/k_{B}T}\right]\nonumber \\
 & \approx1-\ln(h\nu/k_{B}T)\:\:\:\:\:\:\:\:(h\nu\ll k_{B}T)
\end{align}
corresponds to the entropy of a quantum (classical) harmonic oscillator\cite{berens1983}
and $h$ represents the Planck's constant. Entropy associated with
the gas subsystem $S_{g}$ is determined as 
\begin{align}
S_{g} & =Nk_{B}f_{g}\int_{0}^{\infty}F_{g}(\nu)W_{g}d\nu,\label{eq:entropy_gas}
\end{align}
where the weighting function $W_{g}$ is the sum of the ideal gas
(IG) contribution $W_{{\rm IG}}$ and excess (${\rm ex}$) contribution
$W_{{\rm ex}}$,\cite{carnahan1970} i.e. $W_{g}=W_{{\rm IG}}+W_{{\rm ex}}$.
For a given $T$ and $\rho$ (here $\rho=f_{g}N/V$), $W_{{\rm IG}}$
is expressed as

\begin{align}
W_{{\rm IG}} & =\frac{1}{3k_{B}}S_{{\rm IG}}(T,\:\rho)\nonumber \\
 & =\frac{1}{3}\left\{ \frac{5}{2}+\ln\left[\left(\frac{2\pi mk_{B}T}{h^{2}}\right)^{3/2}\frac{V}{f_{g}N}\right]\right\} 
\end{align}
and the corresponding $W_{{\rm ex}}$ should be\cite{oconnell2005,hansen2006}
\begin{align}
W_{{\rm ex}} & =\frac{1}{3k_{B}}S_{{\rm ex}}(T,\:\rho)\nonumber \\
 & =\frac{1}{3}\frac{\gamma\left(3\gamma-4\right)}{\left(1-\gamma\right)^{2}},\label{eq:entropy_excess}
\end{align}
where $S_{{\rm IG}}(T,\,\rho)$ is the entropy per atom of the ideal
gas, $S_{{\rm ex}}$ is the excess entropy per atom of the HS gas,
$\gamma$ ($\gamma\equiv\frac{\pi\rho}{6}\sigma^{3}$, $\sigma$ is
the hard sphere diameter) is the packing fraction. However in the
original 2PT paper, a different weighting function was used,\cite{carnahan1970}

\begin{align}
W_{{\rm ex}}^{\prime} & =\frac{1}{3}\left\{ \frac{\gamma\left(3\gamma-4\right)}{\left(1-\gamma\right)^{2}}+\ln z\right\} ,\label{eq:excess_entropy_lnz}
\end{align}
which in fact corresponds to the excess entropy under identical $T$
and $P$ (see Appendix B for a derivation). In the following session
we show how this extra $\ln z$ term in $W_{{\rm ex}}$ affects the
predicted entropies.

Eqs. (\ref{eq:entropy_solid}) and (\ref{eq:entropy_gas}) contain
integrals with infinity as upper bounds. In practice, this infinite
upper bound is replaced by a finite $\nu_{0}$, which should be sufficiently
large to ensure the numerical integration result converges to the
exact value. This is particularly important for $S_{g}$, where $F_{g}(\nu)$
decreases slowly as $\nu^{-2}$ and $W_{g}$ is $\nu$-independent.
An alternative way to evaluate $S_{g}$ is to apply the sum rule $\int_{0}^{\infty}F_{g}(\nu)d\nu=3$,
which leads to $S_{g}=3Nk_{B}f_{g}W_{g}$. 

\section{liquid argon}

Liquid argon is the prototype of simple fluids. Its phase diagram
and thermodynamic properties have been determined accurately over
a wide $T$ and $\rho$ range.\cite{johnson1993} This makes liquid
argon an ideal model system for methodological developments. The inter-atomic
potential of liquid argon takes the form 
\begin{align*}
V(r) & =4\epsilon\left[\left(\frac{\sigma}{r}\right)^{12}-\left(\frac{\sigma}{r}\right)^{6}\right],
\end{align*}
where $\sigma=3.405$ \AA, $\epsilon/k_{B}=119.8$ K, $m=39.948$
g/mol. For generality, $T$ and $\rho$ are measured in reduced units
as $T^{\star}=k_{B}T/\epsilon$, $\rho^{\star}=\rho\sigma^{3}$. In
the original 2PT paper,\cite{lin2003} Five $\rho^{\star}$ ($1.10$,
$0.85$, $0.70$, $0.40$, and $0.05$) and four $T^{\star}$ ($1.8$,
$1.4$, $1.1$, and $0.9$) were considered. These $(\rho^{\star},\:T^{\star})$
cover fluid, solid, metastable and unstable states. Here we focus
on $(\rho^{\star},\:T^{\star})$ conditions where argon is in the
fluid state, the intended target of the 2PT model. Moreover, the sampling
of $T^{\star}$ is increased to elucidate the trends in $S(T)$. MD
simulations\cite{todorov2006} were performed with the same setup
as Lin et al.,\cite{lin2003} i.e., the system contained 512 atoms,
the time step was 8 fs, each simulation first ran $10\:000$ steps
for equilibration, then another $20\:000$ to $50\:000$ steps for
production. Moreover, the maximum entropy method\cite{press1986}
was applied to minimize the statistical noises and produce smooth
VDoS. 

\begin{figure}
\subfigure[][]{\includegraphics[width=0.42\textwidth]{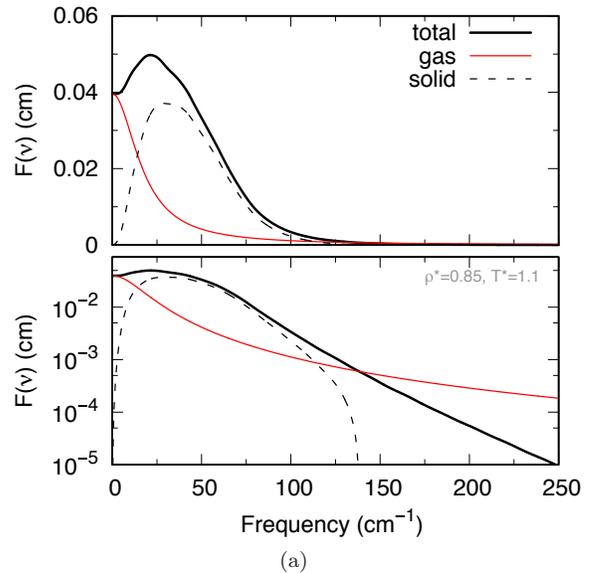}}\hfill{}\subfigure[][]{\includegraphics[width=0.4\textwidth]{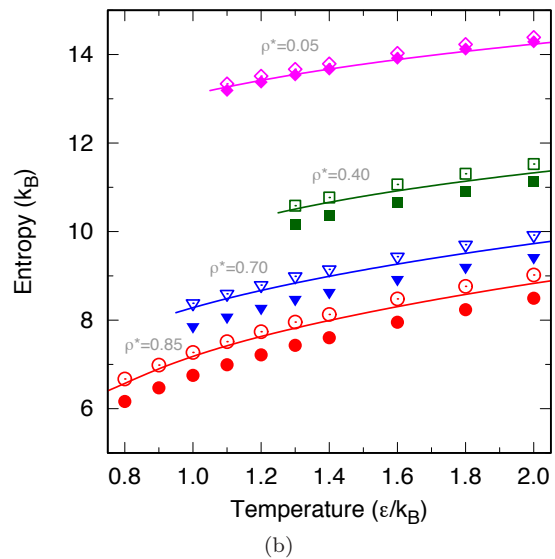}}

\caption{(a) Partitions of the VDoS of liquid argon at $\rho^{\star}=0.85$,
$T^{\star}=1.1$ in the original 2PT model. (b) Entropies at various
$\rho^{\star}$ and $T^{\star}$. Open symbols denote results obtained
with the same $S_{{\rm ex}}$ formula as Lin et al.,\cite{lin2003}
closed symbols denote results with the right formula (without $\ln z$),
solid lines represent the MBWR EOS.\cite{johnson1993}}
\end{figure}

Fig. 1(a) shows a representative $F(\nu)$ at $\rho^{\star}=0.85$,
$T^{\star}=1.1$ and the partition of gas-like and solid-like components.
We see that the gas-like component $f_{g}F_{g}(\nu)$ takes the maximum
at $\nu=0$ where $f_{g}F_{g}(0)=F(0)$, then declines monotonically
as $\nu$ increases. In contrast, the solid-like component $(1-f_{g})F_{s}(\nu)$
is zero at $\nu=0$ but makes predominant contribution to $F(\nu)$
at higher frequencies. A subtle feature, manifest only in the logarithmic
scale, is that $f_{g}F_{g}(\nu)$ becomes larger than the total $F(\nu)$
when $\nu\apprge140$ cm$^{-1}$. This mismatch between $f_{g}F_{g}(\nu)$
and $F(\nu)$ was first noticed by Desjarlais in his AIMD simulations
of liquid metals.\cite{desjarlais2013} Here we see it is also present
in liquid argon. To quantify this mismatch, we define an auxiliary
function $F_{a}(\nu)$ as 
\begin{align*}
F_{a}(\nu) & =\begin{cases}
f_{g}F_{g}(\nu)-F(\nu), & f_{g}F_{g}(\nu)>F(\nu)\\
0, & f_{g}F_{g}(\nu)\le F(\nu)
\end{cases}.
\end{align*}
Although $F_{a}(\nu)$ seems negligibly small, it spans a very wide
frequency range (diminishes as $\nu^{-2}$ at high $\nu$) and the
overall contribution is still notable. At $\rho^{\star}=0.85$, $T^{\star}=1.1$,
the integral $\int_{0}^{\infty}F_{a}(\nu)\:d\nu=0.0625$, or $2\%$
of the total VDoS. Here the integration was performed numerically
by replacing the infinite upper bound with a large $\nu_{0}$ (2084.6
cm$^{-1}$), and the residue $\int_{\nu_{0}}^{\infty}F_{a}(\nu)d\nu$
is less than $5\times10^{-4}$. If we discard the entropy associated
with $F_{a}(\nu)$, either by enforcing the VDoS of the gas-component
equals $F(\nu)$ when $f_{g}F_{g}(\nu)>F(\nu)$, or by adopting a
smaller $\nu_{0}$ (e. g. 200 cm$^{-1}$) when evaluating the entropy
integrals in Eqs. (\ref{eq:entropy_solid}) and (\ref{eq:entropy_gas}),
we are able to reproduce Lin et al.'s results,\cite{lin2003} shown
as open symbols in Fig. 1(b) and tabulated in Table \ref{table:argon}.
These results agree with the modified Benedict-Webb-Rubin (MBWR) EOS\cite{johnson1993}
fairly well, however one should be aware that this agreement is achieved
when (i) the entropy associated with $F_{{\rm a}}(\nu)$ is ignored,
(ii) the excess entropy of HS was evaluated using Eq. (\ref{eq:excess_entropy_lnz}),
instead of Eq. (\ref{eq:entropy_excess}). If the right formula for
weighting function (without $\ln(z)$) was applied, entropy will drop
significantly (solid symbols in Fig. 1(b)). For instance, at $\rho^{\star}=0.85$
, $T^{\star}=1.1$, the original 2PT prediction is $7.51$ $k_{B}$
per atom, about $1$\% higher than the MBWR result ($7.42$ $k_{B}$).
After dropping the $\ln(z)$ term in $W_{g}$, the 2PT prediction
becomes 6.99 $k_{B}$ per atom, nearly $6\%$ lower than the MBWR
result. Only at low densities ( e. g. $\rho^{\star}=0.05$) where
the system is close to ideal gas and $S_{{\rm ex}}$ is small, the
effect of the $\ln(z)$ term is inconsequential. 

\begin{figure}
\includegraphics[width=1.0\columnwidth]{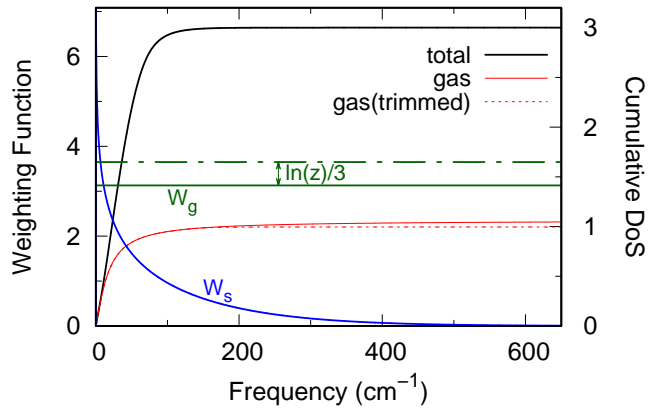}

\caption{Weighting functions (left axis) $W_{s}(\nu)$, $W_{g}$ and cumulative
VDoS (right axis) : $I(\nu)=\int_{0}^{\nu}F(\nu)d\nu$ (``total''),
$I_{g}(\nu)=\int_{0}^{v}f_{g}F_{g}(\nu)d\nu$ (``gas''), and $I_{g,t}(\nu)=\int_{0}^{\nu}[f_{g}F_{g}(\nu)-F_{a}(\nu)]d\nu$
(``gas(trimmed)'') for liquid argon at $\rho^{\star}=0.85$, $T^{\star}=1.1$.}
\end{figure}

To better appreciate these results, we present in Fig. 2 the weighting
functions $W_{s}(\nu)$, $W_{g}$ (with and without $\ln(z)$), and
cumulative VDoS: $I(\nu)=\int_{0}^{\nu}F(\nu)d\nu$ (``total''),
$I_{g}(\nu)=\int_{0}^{v}f_{g}F_{g}(\nu)d\nu$ (``gas''), and $I_{g,t}(\nu)=\int_{0}^{\nu}[f_{g}F_{g}(\nu)-F_{a}(\nu)]d\nu$
(``gas(trimmed)'') at $\rho^{\star}=0.85$ , $T^{\star}=1.1$. The
last integral $I_{g,t}(\nu)$ corresponds to the case where the high-frequency
tail of $f_{g}F_{g}(\nu)$ is trimmed by enforcing the VDoS of the
gas component equals $F(\nu)$ when $f_{g}F_{g}(\nu)>F(\nu)$. At
this $\rho^{\star}$ and $T^{\star}$, $f_{g}=0.35$, $\gamma=0.33$
, and $\ln(z)/3=0.52$ . In comparison, $W_{{\rm ex}}$ and $W_{{\rm IG}}$
equal $-0.75$ and $3.88$, respectively. Therefore with or without
the $\ln(z)$ term in $W_{g}$ makes a big difference in the predicted
2PT entropy. Also, we note that $I(\nu)$ and $I_{g,t}(\nu)$ saturate
with $\nu$ quickly ($\sim200$ cm$^{-1}$) , while $I_{g}(\nu)$
saturates much slowly due to the presence of $F_{a}(\nu)$. If we
include $F_{a}(\nu)$ in the VDoS of the gas component, $S_{g}$ will
increase by $\int_{0}^{\infty}F_{a}(\nu)W_{g}d\nu=0.20$ $k_{B}$
per atom; if we treat $F_{a}(\nu)$ as harmonic oscillators (solid-component),
the corresponding entropy is $\int_{0}^{\infty}F_{a}(\nu)W_{s}(\nu)d\nu=0.01$
$k_{B}$ per atom. The difference between these two cases is because
in regions where $F_{a}(\nu)$ is non-zero, $W_{s}(\nu)\ll W_{g}$.
These information will be useful when we try to revise the 2PT model.

\section{revised 2PT model}

\begin{table*}
\caption{Entropies (in $k_B$ per atom) of Lennard-Jones fluid at selected $\rho^{\star}$ and $T^{\star}$. ``MBWR EOS'' denotes data from Ref. \onlinecite{johnson1993}, ``2PT'' and ``2PT (w/o $\ln z$)'' correspond to results from the original 2PT formalism with and without the $\ln(z)$ term in $W_g$, ``R2PT ($\delta$)'' represents results from the revised 2PT model with the value of $\delta$ in the parentheses. Uncertainties in the 2PT calculations are estimated to be $0.02$ $k_B$ per atom.}
\label{table:argon}
\begin{ruledtabular}

\begin{tabular*}{1\textwidth}{@{\extracolsep{\fill}}ccccccc}
$\rho^{\star}$ & $T^{\star}$ & MBWR EOS & 2PT & 2PT (w/o $\ln z$) & R2PT ($1.0$) & R2PT ($1.5$)\tabularnewline
\midrule
 & 0.8 & 6.57 & 6.67 & 6.16 & 6.33 & 6.50\tabularnewline
 & 1.1 & 7.42 & 7.51 & 6.99 & 7.18 & 7.36\tabularnewline
0.85 & 1.4 & 7.99 & 8.13 & 7.60 & 7.79 & 7.98\tabularnewline
 & 1.8 & 8.58 & 8.76 & 8.23 & 8.43 & 8.62\tabularnewline
 & 2.0 & 8.83 & 9.02 & 8.49 & 8.70 & 8.88\tabularnewline
 &  &  &  &  &  & \tabularnewline
 & 1.0 & 8.28 & 8.38 & 7.86 & 8.04 & 8.21\tabularnewline
0.70 & 1.4 & 8.99 & 9.15 & 8.63 & 8.83 & 8.99\tabularnewline
 & 1.8 & 9.51 & 9.70 & 9.20 & 9.40 & 9.55\tabularnewline
 &  &  &  &  &  & \tabularnewline
 & 1.3 & 10.51 & 10.58 & 10.17 & 10.33 & 10.39\tabularnewline
0.40 & 1.6 & 10.92 & 11.06 & 10.67 & 10.83 & 10.88\tabularnewline
 & 1.8 & 11.14 & 11.31 & 10.91 & 11.08 & 11.13\tabularnewline
 &  &  &  &  &  & \tabularnewline
 & 1.1 & 13.27 & 13.34 & 13.18 & 13.24 & 13.19\tabularnewline
0.05 & 1.4 & 13.67 & 13.79 & 13.67 & 13.72 & 13.67\tabularnewline
 & 1.8 & 14.07 & 14.23 & 14.11 & 14.17 & 14.12\tabularnewline
\end{tabular*}

\end{ruledtabular}
\end{table*}

In the prior section we demonstrate a dilemma in the original 2PT
formalism: the spurious $\ln(z)$ term in evaluating $S_{{\rm ex}}$
should be dropped, but removing this term spoils the good agreement
and entropy becomes significantly underestimated. A question arises
naturally: is it possible to modify the 2PT formalism, such that it
adopts the right formula for $S_{{\rm ex}}$ and yet gives accurate
entropy? In the following we construct a revised 2PT model that meets
this requirement.

We have shown that $f_{g}F_{g}(\nu)$ is larger than $F(\nu)$ at
high frequencies and the excess VDoS is defined as $F_{a}(\nu)$.
Since entropies associated with $F_{a}(\nu)$ were not accounted for
in the original calculation, our first revision of the model is to
include such contributions. This is more sensible than simply discarding
$F_{a}(\nu)$ if one considers the fundamental difference between
diffusion and vibration: For vibrations, it is okay to consider entropy
contributions from various frequency components of VDoS individually,
as they represent different harmonic oscillators whose motions are
independent. Diffusion is a far more complex type of motion with all
frequency components coupled to each other and only $f_{g}F_{g}(\nu)$
as a whole is physically meaningful. To determine entropy associated
with $f_{g}F_{g}(\nu)$, we replace the numerical integration $S_{g}=Nk_{B}f_{g}\int_{0}^{\nu_{0}}F_{g}(\nu)W_{g}d\nu,$
where $\nu_{0}$ is the upper bound of the VDoS, with the closed form
$3Nk_{B}f_{g}W_{g}$. This allows us to avoid the truncation error
that may arise from a finite $\nu_{0}$. A consequence of including
$F_{a}(\nu)$ in the VDoS of the gas component is that the corresponding
VDoS of the solid component becomes negative, as $(1-f_{g})F_{s}(\nu)\equiv F(\nu)-f_{g}F_{g}(\nu)=-F_{a}(\nu)$.
The appearance of negative solid VDoS can be interpreted as follows:
Imagine the liquid system under study is attached with an assemblage
of auxiliary harmonic oscillators with VDoS $F_{a}(\nu)$. The VDoS
of the combined system is $F(\nu)+F_{a}(\nu)$. To get entropy associated
with $F(\nu)$, one may first perform standard 2PT calculation on
the combined system, then subtract the entropies of auxiliary harmonic
oscillators. Thus the negative solid VDoS is just book-keeping of
auxiliary harmonic oscillators, with entropy $S_{s,a}=-Nk_{B}\int_{0}^{\infty}F_{a}(\nu)W_{s}(\nu)d\nu$. Note at high $T$, $f_{g}F_{g}(\nu)$ may 
slightly exceed $F(\nu)$ also near $\nu=0$ (see supplementary material). 
Such mismatch is readily handled via $F_{a}(\nu)$, just like the mismatch 
in the high $\nu$ region.
In contrast to $S_{g}$, $S_{s,a}$ is easy to converge in numerical
integration as $W_{s}(\nu)$ decreases exponentially at high $\nu$.
The value of $S_{s,a}$ is small. At $\rho^{\star}=0.85$, $T^{\star}=1.1$,
$S_{s,a}$ is just $-0.01$ $k_{B}$ per atom, whereas the increase
in $S_{g}$ after accounting for $F_{a}(\nu)$ is $0.20$ $k_{B}$
per atom. Thus the overall effect of including $F_{a}(\nu)$ is to
increase entropy by $0.19$ $k_{B}$ per atom. This moves the 2PT
predictions (tabulated with label ``R2PT(1.0)'' in Table \ref{table:argon})
closer to the standard MBWR results, but the remaining differences
($3\%$) is still substantial compared to the desired accuracy ($1\%$)
in practical applications.\cite{fultz2010} 

\begin{figure}
\includegraphics[width=1.0\columnwidth]{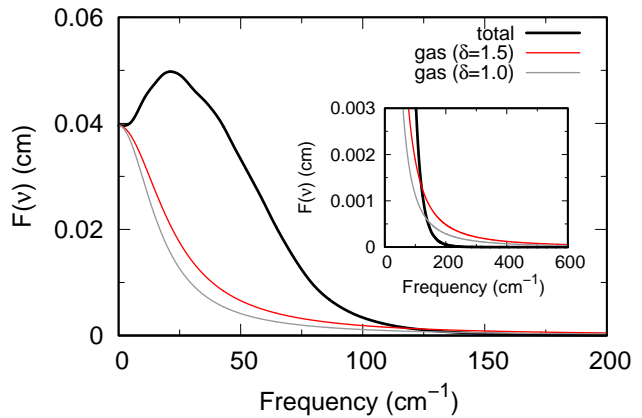}

\caption{Partitions of the VDoS of liquid argon at $\rho^{\star}=0.85$, $T^{\star}=1.1$
with the gas fraction $f_{g}$ determined from $\delta=1.5$ and $1.0$,
respectively. Inset: amplified VDoS demonstrating the high frequency
tail $F_{a}(\nu)$.}
\end{figure}

\begin{figure}
\includegraphics[width=0.85\columnwidth]{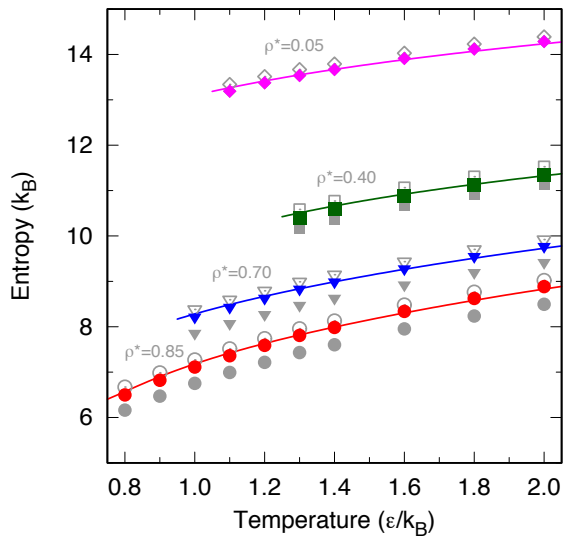}

\caption{Entropies of liquid argon at various $\rho^{\star}$ and $T^{\star}$.
Closed symbols represent predictions from the revised 2PT model with
$\delta=1.5$, solid lines are MBWR results, grey open (filled) symbols
stand for entropies from the original 2PT model with (without) the
$\ln(z)$ term.}
\end{figure}

To further improve the accuracy of the model, we direct our attention
to the gas fraction $f_{g}$, a central parameter in the 2PT model
dictating the partition of gas-solid components. As detailed in Appendix
A, $f_{g}$ is determined from two assumptions whose theoretical footings
are not equal. One is based on the classic Enskog theory of dense
HS, while the other, $f_{g}=D/D_{0}$ with $D_{0}$ the diffusivity
of dilute HS, is more speculative. It is made under the consideration
that $f_{g}$ should be $0$ when the diffusivity of the system is
$0$ and approach $1$ in the high temperature-low density limit.
Yet this consideration can be served equally well by assuming $f_{g}^{\delta}=D/D_{0}$,
where the exponent $\delta$ is not restricted to unity. We emphasize
that the original $\delta=1$ may seem natural, but in essence it
is a tacit assumption made by Lin et al. and is not required by any
physical law. Indeed, if one considers $\tilde{f}_{g}^{\delta}=D/D_{0}$
as the definition of $\delta$, with $\tilde{f}_{g}$ the optimal
$f_{g}$ that yields the exact entropy of the system, then $\delta$
will vary from system to system. When the variation of $\delta$ is
small, it can be approximated as a constant. At present, we lack physical
constraints to compute $\delta$ directly, so its optimal value is
determined by comparing with outside references. 

After a few tests, we find $\delta=1.5$ a suitable choice for liquid
Ar. Partitions of VDoS at $\rho^{\star}=0.85$, $T^{\star}=1.1$ using
$\delta=1.5$ as well as the original $\mbox{\ensuremath{\delta}=}1.0$
are shown in Fig. 3. The adoption of $\delta=1.5$ increases $f_{g}$
from $0.35$ to $0.46$ and the packing fraction $\gamma$ from $0.33$
to $0.36$. The corresponding entropy increases from $7.18$ to $7.36$
$k_{B}$ per atom, within $1\%$ of the MBWR value ($7.42$ $k_{B}$).
Entropies at other $\rho^{\star}$-$T^{\star}$ conditions are shown
in Fig. 4 and tabulated in Table \ref{table:argon} with label ``R2PT(1.5)''.
We see that the overall agreement with respect to MBWR is quite good,
even slightly better than the original 2PT predictions (with the $\ln(z)$
term). 

\section{liquid metals}

We now consider an important yet more complex class of fluids: liquid
metals. They have broad applications in industry\cite{scopigno2005}
as well as in Earth sciences.\cite{belonoshko2000} Deploying the
2PT model to liquid metals was initiated by Desjarlais,\cite{desjarlais2013}
who proposed a novel memory-function formalism to resolve the perceived
overestimation of entropy due to the high frequency tail of $f_{g}F_{g}(\nu)$.
However as we noted before, contributions from high frequency tails
were not included in Lin et al.'s calculations in the first place,
so the actual difference between entropies predicted from the two
approaches is minor. Moreover, both approaches
rely on the $\ln(z)$ term in the weighting function to produce good
results. As we now have a revised 2PT model that works well for liquid
Ar, it is tempting to extend this model to liquid metals.

\begin{table*}
\caption{Ionic entropies (in $k_B$ per atom) of various metals near the experimental melting temperature. Here the density $\rho$ is in g/cm$^3$, temperature $T$ is in Kelvin. ``TI'' stands for thermodynamic integration. ``R2PT ($\delta$)'' for the revised 2PT formalism, with the value of $\delta$ in the parentheses. ``2PT'' and ``2PT (w/o $\ln z$)'' correspond to results from the original 2PT formalism with and without the $\ln(z)$ term in $W_g$. ``($n$,\, $m$)'' denotes the exponents of the Sutton-Chen potential. Uncertainties in the TI and 2PT calculations are estimated to be $0.01$ and $0.02$ $k_B$ per atom, respectively.}
\label{table:metal}
\begin{ruledtabular}

\begin{tabular*}{1\textwidth}{@{\extracolsep{\fill}}ccccccccc}
 & $\rho$ & $T$ & TI & R2PT ($1.5$) & R2PT ($1.0$) & 2PT & 2PT (w/o $\ln z$) & $(n,\:m)$\tabularnewline
\midrule
Ir & 19.0 & 2700 & 12.25 & 12.19 & 11.99 & 12.28 & 11.80 & $(14,\:6)$\tabularnewline
Ag & 9.32 & 1200 & 10.24 & 10.25 & 10.06 & 10.33 & 9.86 & $(12,\:6)$\tabularnewline
Rh & 10.7 & 2200 & 11.10 & 11.09 & 10.89 & 11.17 & 10.68 & $(12,\:6)$\tabularnewline
Pd & 10.38 & 1800 & 11.08 & 11.05 & 10.84 & 11.13 & 10.64 & $(12,\:7)$\tabularnewline
Au & 17.31 & 1300 & 11.71 & 11.73 & 11.52 & 11.80 & 11.29 & $(10,\:8)$\tabularnewline
Ni & 7.81 & 1700 & 10.31 & 10.38 & 10.17 & 10.43 & 9.93 & $(9,\:6)$\tabularnewline
Al & 2.375 & 900 & 9.25 & 9.40 & 9.20 & 9.46 & 8.94 & $(7,\:6)$\tabularnewline
\end{tabular*}

\end{ruledtabular}
\end{table*}

\begin{figure}
\includegraphics[width=1.0\columnwidth]{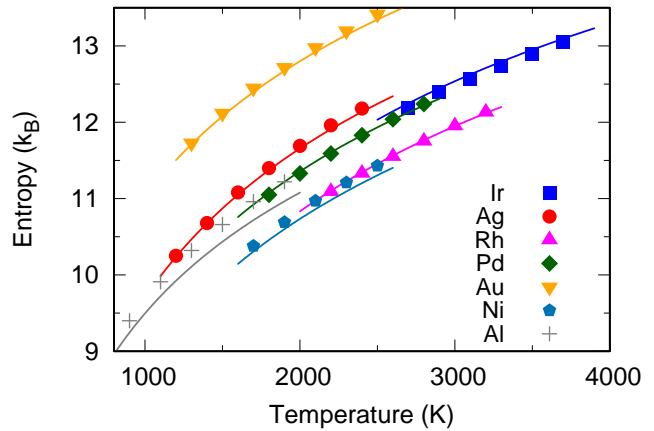}

\caption{Ionic entropies of liquid metals using the Sutton-Chen potential.
Solid lines are references from TI, points are from the revised 2PT
model with $\delta=1.5$. }
\end{figure}

We choose the well-established Sutton-Chen many-atom potential\cite{sutton1990}
to describe the interatomic interactions in liquid metals. With the
Sutton-Chen potential, the total potential energy of the system takes
the form 
\begin{align}
U & =\epsilon\left[\frac{1}{2}\sum_{i}\sum_{j\ne i}V(r_{ij})-c\sum_{i}\sqrt{\rho_{i}}\right],
\end{align}
where $V(r)=(a/r)^{n}$, $\rho_{i}=\sum_{j\ne i}(a/r_{ij})^{m}$.
Among them, $\epsilon$ is a parameter with the dimension of energy,
$r_{ij}$ is the separation between atoms $i$ and $j$, $c$ is a
positive dimensionless parameter, $a$ is a parameter with the dimension
of length, $n$ and $m$ are positive integers. In some respects,
the Sutton-Chen potential resembles the Lennard-Jones potential for
liquid argon, with $V(r)$ corresponding to the $(\sigma/r)^{12}$
repulsive part and $-c\sqrt{\rho_{i}}$ to the $-(\sigma/r)^{6}$
attractive part. The pair $(n,\:m)$ determines how the potential
varies with interatomic distances and serves as a measure of the softness
of potentials. To see how our revised model performs with different
potentials, we choose seven metals with $(n,\:m)$ ranging from $(14,\:6)$
as Ir, to $(7,\:6)$ as Al.\cite{sutton1990} For each metal, we fix
its density to the experimental density at melting point $T_{m}$,
with temperature ranging from close to $T_{m}$ to $1000$ K above
$T_{m}$. The $NVT$ MD simulations\cite{todorov2006} were performed
in a cubic cell containing $512$ atoms. The time step was $1$ fs.
Each simulation first ran $50$ ps for thermal equilibration, then
another $100$ ps for production. Calculations of the 2PT model were
conducted in the same fashion as in liquid argon. To measure the accuracy
of 2PT results, we further performed extensive TI calculations (see
Appendix C for details) from which we extract entropies as references.

\begin{figure*}
\subfigure[][]{\includegraphics[width=0.5\textwidth]{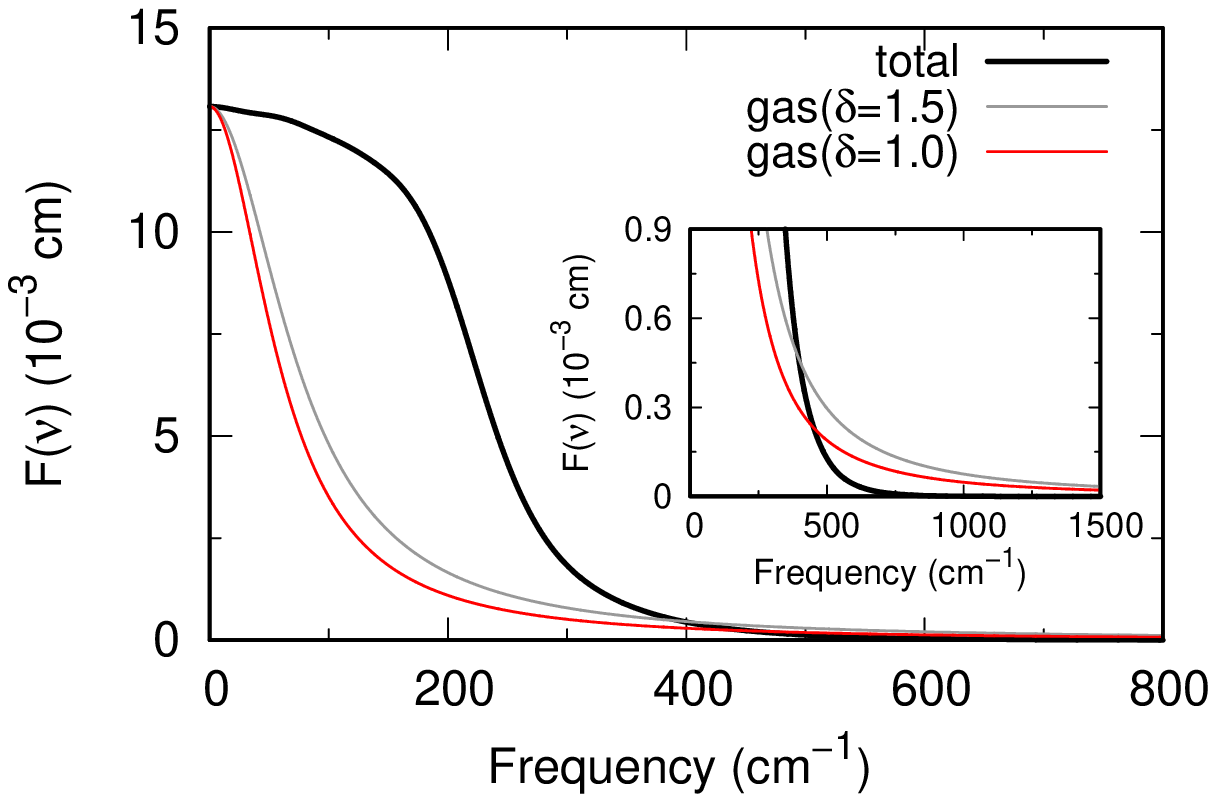}}\hfill{}\subfigure[][]{\includegraphics[width=0.5\textwidth]{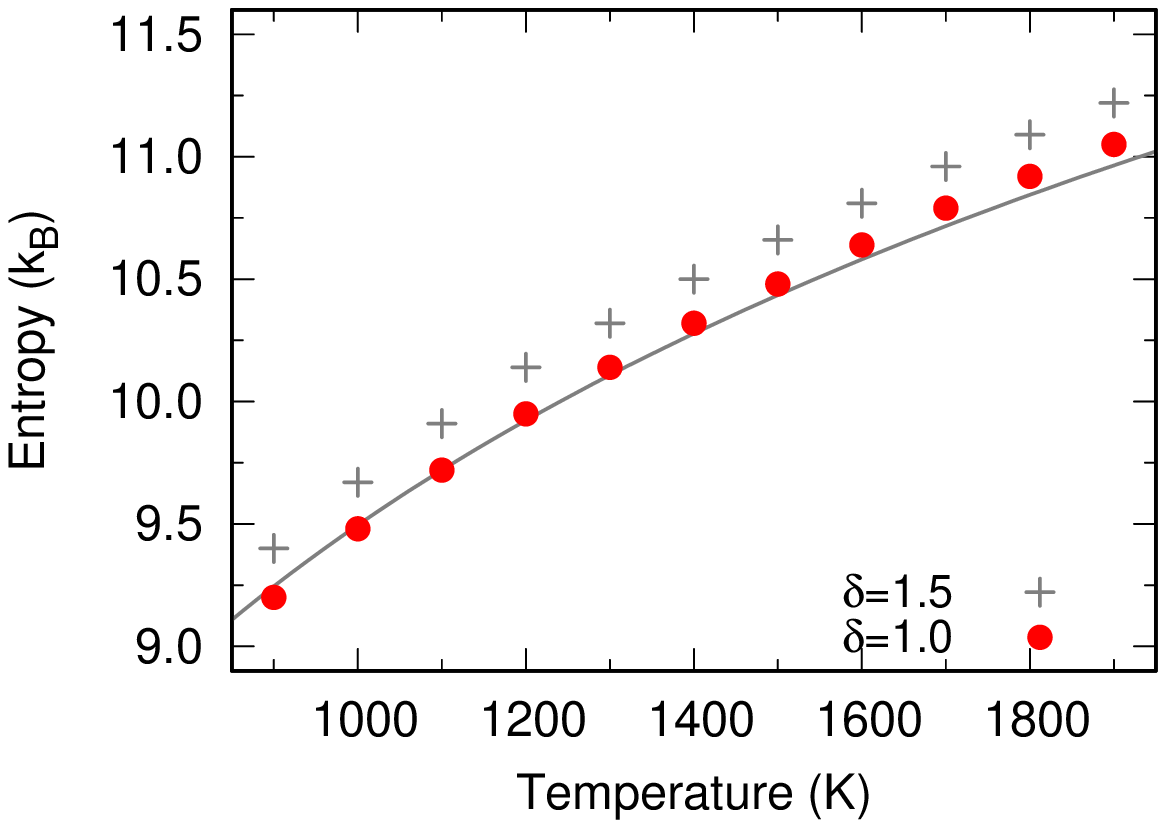}}

\caption{(a) Partitions of VDoS with $\delta=1.5$ and $1.0$ for liquid Al
at $\rho=2.375$ g/cm$^{3}$ and $T=900$ K. Inset: amplified VDoS
demonstrating the high frequency tail $F_{a}(\nu)$. (b) Ionic entropies
of Al at $\rho=2.375$ g/cm$^{3}$. Solid lines are references from
TI, points are predictions from the revised 2PT model with $\delta=1.5$
and $1.0$, respectively.}
\end{figure*}

Figure 5 compares entropies determined from our revised 2PT model
using $\delta=1.5$ with those from TI. The corresponding data are
tabulated in Table \ref{table:metal}. We see that the overall agreement
is good, especially in Ag, Rh, Pd and Au. Still, noticeable discrepancies
appear in Ir and Ni, with entropy of the former being underestimated
and the latter being overestimated. More significant overestimations
are seen in Al. To understand this phenomenon, recall that our revised
2PT model was initially developed for liquid Ar, where the exponent
$n\:(m)$ for the repulsive (attractive) part of potential is $12\:(6)$.
Apparently, our model works best for liquid metals with $n=12$, such
as Ag, Rh, Pd, whereas it underestimates entropy for liquid metals
with harder ($n>12)$ potentials and overestimates for those with
softer $(n<12)$ potentials. The largest overestimation takes place
in Al whose potential is the softest $(n=7)$. In this worst case,
the error is $0.15$ $k_{B}$ per atom at $900$ K, or $1.6\%$ of
the total ionic entropy. To put this level of accuracy into context,
we note the absolute (relative) error at $\rho^{\star}=0.85$, $T^{\star}=0.9$
(near the triple point) is $0.1$ $k_{B}$ ($1.5\%$ ) when the original
2PT model is applied to the Lennard-Jones system.\cite{lin2003} 
With $\delta=1.5$
as the default value, our revised 2PT formalism is suited for quick
estimations of entropies for a wide range of materials at various
conditions, with accuracy comparable to other 2PT formalisms\cite{lin2003,desjarlais2013,meyer2016}
proposed before.

Further improvements in accuracy can be achieved if one does not
require a fixed $\delta$ for all materials. From the study of liquid
argon, we know that the 2PT entropy relies on the gas-solid partition,
and we have introduced a parameter $\delta$ to refine this partition.
For Ar, setting $\delta=1.0$ would underestimate entropy while the
optimal $\delta=1.5$. This indicates that one may accommodate potentials
of different nature by adjusting the parameter $\delta.$ Indeed,
take liquid Al as an example, changing $\delta$ from the default
value of $1.5$ to $1.0$ yields a smaller gas fraction $f_{g}$,
as shown in Fig. 6(a), the resulting entropy also decreases and the
agreement with TI gets significantly improved, as shown in Fig. 6(b). 

In some situations, quick estimations may not be good enough and one
would like to determine the absolute entropy as precise as possible.
In principle, TI is the method of choice for such cases. However the
computational cost to perform TI can be prohibitively high for AIMD
of large systems. In such cases, our revised 2PT model may serve as
a viable alternative: one may first perform both the 2PT and
TI calculations in a small supercell with fewer atoms at the targeted
$T$ and $\rho$, identify the best $\delta$ for this particular
thermal state, then apply this $\delta$ to larger supercells. Good
transferability of $\delta$ from small to large cells is anticipated
since $\delta$ is controlled by interatomic interactions and should
not depend strongly on supercell size. This feature can be quite useful
for accurate determination of chemical potentials, phase boundaries,
etc., for large and complex systems.

\section{conclusion}

We have conducted a detailed analysis on the 2PT model for computing
the ionic entropy of liquids in MD.  This analysis reaffirms the seminal 
idea of Lin, Blanco, and Goddard,\cite{lin2003} namely, thermodynamic 
properties of liquids can be accurately determined by decomposition of 
VDoS using two idealized 
systems of hard-sphere gas and harmonic oscillators.  Some deficiencies 
in the original model are also identified and addressed. In particular,
the spurious $\ln(z)$ term is removed
from the weighting function; The correspondence between dynamics and
thermodynamics is enforced by considering the VDoS of the gas component
as a whole when determining its associated entropy; Partition of gas-solid
components is now subject to optimization for better accuracy. With
improved theoretical formality, the new formalism is ready to be applied
to a wide range of systems for quick entropy estimations. Moreover,
it can be combined with TI to obtain accurate entropies of specific
thermal states. This latter feature will be useful in situations where
high accuracy is necessary, but the systems are too large to perform
TI directly. 

\section*{supplementary material}
See supplementary material on the mismatch of HS and total VDoS near 
$\nu = 0$ at high $T$.

\begin{acknowledgments}
We thank M. P. Desjarlais for stimulating discussion. This work is
supported by Ministry of Science and Technology of China grant No.
2014CB845905, the Strategic Priority Research Program (B) of the Chinese
Academy of Sciences (XDB18000000), National Natural Science Foundation
of China grant 41474069, Special Program for Applied Research on Super 
Computation of the NSFC-Guangdong Joint Fund (the second phase) under 
grant No. U1501501 and Computer Simulation Lab, IGGCAS. Calculations 
were performed on TianHe-1A supercomputer at the National Supercomputer 
Center of China(NSCC) in Tianjin.
\end{acknowledgments}

\appendix

\section{determination of $f_{g}$}

A central parameter in the 2PT model is the gas fraction $f_{g}$
. Equations to determine $f_{g}$ were introduced by Lin et al..\cite{lin2003}
Here we slightly generalize Lin et al.'s approach to improve its applicability. 

First, consider that $f_{g}$ should be $0$ when the diffusivity
of the system is $0$ (the system is completely solid with no gas
fraction) and approach $1$ in the high temperature-low density limit
(the system becomes completely gas), thus $f_{g}$ can be defined
as 
\begin{align}
f_{g}^{\delta} & =\frac{D}{D_{0}(N)},\label{eq:fg_D0_D}
\end{align}
where $D$ is the diffusivity of the MD system, determined from $F(0)$
using Eq. (\ref{eq: f0 and d}), $D_{0}(N)$ is the diffusivity of
HS gas in the low density limit (the Chapman-Enskog result),\cite{mcquarrie2000,hansen2006}
defined as 
\begin{align}
D_{0}(N) & =\frac{3}{8}\frac{V}{N\sigma^{2}}\left(\frac{k_{B}T}{\pi m}\right)^{1/2},
\end{align}
with $N$ the total number of particles in the MD system, $V$ the
system volume, $m$ the mass of the particle, $\sigma$ the HS diameter,
$k_{B}$ the Boltzmann constant. The exponent $\delta$ was set to
unity in the original Lin et al.'s paper.\cite{lin2003} Here we allow
it to be system-dependent. Define a normalized diffusivity $\Delta$
as
\begin{align}
\Delta & \equiv\frac{8}{3}(\frac{36}{\pi^{2}})^{1/3}D\sqrt{\frac{\pi m}{k_{B}T}}\left(\frac{N}{V}\right)^{1/3}
\end{align}
and packing fraction $\gamma\equiv\frac{\pi f_{g}N}{6V}\sigma^{3}$,
Eq. (\ref{eq:fg_D0_D}) can be simplified as
\begin{align}
\gamma & =\Delta^{-\frac{3}{2}}f_{g}^{\frac{3}{2}\delta+1}.\label{eq:gamma}
\end{align}

The second consideration is that in the spirit of 2PT model, the diffusivity
$D$ is solely caused by motions of $f_{g}N$ particles. The rest
$(1-f_{g})N$ particles just vibrate and do not diffuse (no contribution
to $F(0)$). As such, 
\begin{align}
D_{{\rm HS}}(f_{g}N) & =\frac{D}{f_{g}}.\label{eq:fg_Dhs_D}
\end{align}
Here $D_{{\rm HS}}(f_{g}N)$ is the diffusivity of HS gas with $f_{g}N$
particles in volume $V$. The $1/f_{g}$ factor on the right-hand
side reflects that $D$ is an apparent diffusivity determined from
the VACF of $N$ particles, whereas the gas subsystem only contains
$f_{g}N$ particles and its actual diffusivity is $1/f_{g}$ times
higher than $D$. An analytic theory for $D_{{\rm HS}}$ was given
by Enskog as 
\begin{align}
D_{{\rm HS}}(f_{g}N) & =D_{0}(f_{g}N)/g(\sigma^{+}),
\end{align}
where $g(\sigma^{+})$ is the value of radial distribution function
at contact.\cite{hansen2006} With Carnahan-Starling equation of state,
$g(\sigma^{+})=(z-1)/4\gamma$, where the compressibility $z=(1+\gamma+\gamma^{2}-\gamma^{3})/(1-\gamma)^{3}$.\cite{hansen2006}
Eq. (\ref{eq:fg_Dhs_D}) can be simplified as
\begin{align}
f_{g}^{2/3} & =\frac{\Delta}{2}\frac{\gamma^{2/3}(2-\gamma)}{(1-\gamma)^{3}}.\label{eq:fg}
\end{align}
Substituting $\gamma$ in Eq. (\ref{eq:fg}) using Eq. (\ref{eq:gamma}),
one reaches the equation for $f_{g}$ as

\begin{align}
\Delta^{-\frac{9}{2}}f_{g}^{(3+\frac{9\delta}{2})}-3\Delta^{-3}f_{g}^{(3\delta+2)}+3 & \Delta^{-\frac{3}{2}}f_{g}^{(1+\frac{3}{2}\delta)}\nonumber \\
-\frac{1}{2}\Delta^{-\frac{3}{2}}f_{g}^{(1+\frac{5}{2}\delta)}+f_{g}^{\delta}-1=0.
\end{align}
For $\delta=1$, the above equation becomes Eq. (34) in Ref. {[}\onlinecite{lin2003}{]}.

\section{excess entropy of hard sphere gas}

The excess entropy $S_{{\rm ex}}$ is defined as the entropy difference
between non-ideal and ideal gases under the same physical conditions,
either identical $T$ and $\rho$ or identical $T$ and $P$.\cite{oconnell2005}
We first consider identical $T$ and $\rho$, for which 
\begin{align}
S_{{\rm ex}}(T,\:\rho) & =S_{{\rm HS}}(T,\:\rho)-S_{{\rm IG}}(T,\:\rho).
\end{align}
$S_{{\rm IG}}(T,\:\rho)$ is described by the Sackur-Tetrode formula
as 
\begin{align}
S_{{\rm IG}}(T,\:\rho) & =k_{B}[\frac{5}{2}+\frac{3}{2}\ln(\frac{2\pi mk_{B}T}{h^{2}})-\ln(\rho)],
\end{align}
whereas $S_{{\rm ex}}(T,\:\rho)$ is readily determined as follows:
consider an isotherm from the present $\rho$ to $\rho\rightarrow0$
limit as the thermodynamic integration path, we have 
\begin{align}
S_{{\rm ex}}(T,\:\rho) & =\int_{0}^{\rho}\frac{d\rho}{\rho^{2}}\left(\beta_{V}^{{\rm IG}}-\beta_{V}^{{\rm HS}}\right),
\end{align}
where $\beta_{V}=(\partial S/\partial V)_{T}=(\partial P/\partial T)_{V}$
is the thermal pressure coefficient. Note as $\rho$ decreases, non-ideal
gas becomes increasingly similar to ideal gas, and $S_{{\rm HS}}$
equals $S_{{\rm IG}}$ at the $\rho\rightarrow0$ limit.

From equation of state $P/\rho k_{B}T=z$, where $z$ is the compressibility
($z=1$ for ideal gas), one finds $\beta_{V}^{{\rm HS}}=z\rho k_{B}$
and $\beta_{V}^{{\rm IG}}=\rho k_{B}$. With Carnahan-Starling equation
of state\cite{carnahan1970} $z=(1+\gamma+\gamma^{2}-\gamma^{3})/(1-\gamma)^{3}$
and packing fraction $\gamma=\frac{\pi}{6}\rho\sigma^{3}$, $S_{{\rm ex}}$
is evaluated as
\begin{align}
S_{{\rm ex}}(T,\:\rho) & =k_{B}\int_{0}^{\rho}\frac{d\rho}{\rho}\left(1-z\right)\nonumber \\
 & =k_{B}\frac{3\gamma^{2}-4\gamma}{(1-\gamma)^{2}}.
\end{align}
To get excess entropy under identical $T$ and $P$, we note that
an IG system with density $z\rho$ has the same $P$ as an HS system
with density $\rho$. Accordingly, 
\begin{align}
S_{{\rm IG}}(T,\:P) & =S_{{\rm IG}}(T,\:z\rho)\nonumber \\
 & =k_{B}\left[\frac{5}{2}+\frac{3}{2}\ln(\frac{2\pi mk_{B}T}{h^{2}})-\ln(\rho)-\ln z\right].
\end{align}
And the excess entropy 
\begin{align}
S_{{\rm ex}}(T,\:P) & =S_{{\rm ex}}(T,\:\rho)+k_{B}\ln z.
\end{align}

\section{thermodynamic integration}

\begin{figure}
\label{fig:TI}\includegraphics[width=1.0\columnwidth]{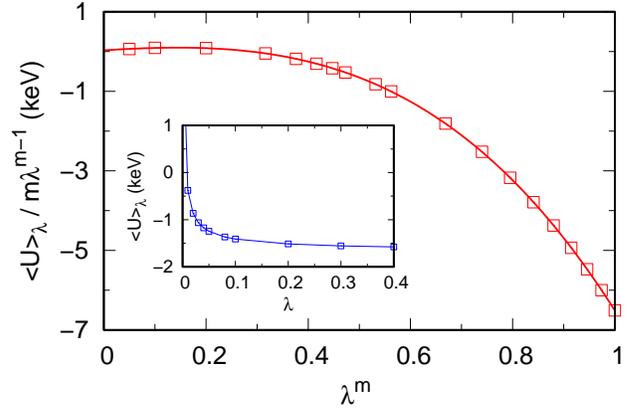}

\caption{Integrand of TI after transforming $\lambda$ to $\lambda^{m}$ $(m=0.25)$.
Squares are data from MD, solid line corresponds to numerical fitting
to the fourth order polynomial. Inset: the original integrand $\langle U\rangle_{\lambda}$,
which diverges when $\lambda\rightarrow0$. The target state is liquid
Al at $\rho=2.375$ g/cm$^{3}$, $T=1000$ K.}
\end{figure}

Thermodynamic integration (TI) is a formally exact method to determine
entropy $(S)$ or free energy $(F)$ of a target state. To perform
TI, one first choose a reference state whose thermodynamic properties
are known; then, a continuous transition path is constructed to connect
the reference and target state; finally, $F$ of the target state
is obtained by combining that of the reference with the free energy
change along the transition path. For a liquid at $(T_{0},\:\rho_{0})$,
the reference state is usually set to be ideal gas at $(T_{0},\:\rho_{0})$.
The transition path consists of states whose interatomic interactions
are gradually switched off.\cite{mezei1989} Denote the potential
energy of the liquid as $U$, states along the transition path as
$U_{\lambda}\equiv\lambda U$, where $\lambda$ is the integration
variable ranging from $0$ to $1$, we have
\begin{align}
\frac{\partial F}{\partial\lambda} & =\langle\frac{\partial U_{\lambda}}{\partial\lambda}\rangle_{\lambda}\nonumber \\
 & =\langle U\rangle_{\lambda},
\end{align}
where $\langle\rangle_{\lambda}$ denotes ensemble average with $U_{\lambda}$
being the potential energy of the ensemble. $F$ of the target state
is then determined as
\begin{align}
F & =F_{{\rm IG}}+\int_{0}^{1}\langle U\rangle_{\lambda}d\lambda,
\end{align}
where $F_{{\rm IG}}$ is the free energy of ideal gas at $(T_{0},\:\rho_{0})$.
This integral is ill-defined because $\langle U\rangle_{\lambda}\rightarrow\infty$
when $\lambda\rightarrow0$. To remove this divergence at $\lambda=0$,
a common practice\cite{mruzik1976,mezei1989} is to transform $\lambda$
to $\lambda^{m}$, such that

\begin{align}
F & =F_{{\rm IG}}+\frac{1}{m}\int_{0}^{1}\frac{\langle U\rangle_{\lambda}}{\lambda^{m-1}}d\lambda^{m}.
\end{align}
Following Ref. {[}\onlinecite{mruzik1976}{]} we set $m=0.25$. As
shown in Fig. 7, the new integrand is well defined at
$\lambda=0$ and the integral can be easily evaluated numerically.
Once $F$ is known, $S$($T_{0},\:\rho_{0})$ is evaluated as $S=\left(E-F\right)/T_{0}$.
Entropy at other temperatures is determined as

\begin{align}
S(T,\:\rho_{0}) & =S(T_{0},\:\rho_{0})+\int_{T_{0}}^{T}\frac{C_{V}(T^{\prime},\:\rho_{0})}{T^{\prime}}dT^{\prime},
\end{align}
where $C_{V}=(\partial E/\partial T)_{V}$ is the heat capacity at
constant volume and can be evaluated from a series of MD simulations
with density fixed at $\rho_{0}$ and temperatures sampled between
$T_{0}$ and $T$.\cite{sun2010,sun2014}

%\bibliographystyle{apsrev4-1}
%\bibliography{tpt}

\end{document}